\documentclass[12pt]{iopart}
\usepackage{color}
\usepackage{epsfig}
\usepackage{psfrag}
\usepackage{graphicx}
\usepackage{color}

\begin{document}

\title{Stability of Fractional Quantum Hall States in Disordered Photonic Systems}

\author{Wade DeGottardi$^1$ and Mohammad Hafezi$^2$}

\address{$^1$Institute for the Research in Electronics and Applied Physics, University of Maryland, College Park, Maryland 20742, USA}

\address{$^2$Joint Quantum Institute, University of Maryland, College Park, Maryland, 20742, USA}

\date{\today}

\begin{abstract}
The possibility of realizing fractional quantum Hall liquids in photonic systems has attracted a great deal of interest of late. Unlike electronic systems, interactions in photonic systems must be engineered from non-linear elements and are thus subject to positional disorder. The stability of the topological liquid relies on repulsive interactions. In this paper we investigate the stability of fractional quantum Hall liquids to impurities which host attractive interactions. We find that for sufficiently strong attractive interactions these impurities can destroy the topological liquid. However, we find that the liquid is quite robust to these defects, a fact which bodes well for the realization of topological quantum Hall liquids in photonic systems.
\end{abstract}


\maketitle

\section{Introduction}

Gapped two-dimensional quantum systems can be classified according their topological properties~\cite{tknn,kane}. Topologically non-trivial systems, which encompasses topological band insulators and topologically ordered systems, exhibit a number of desirable properties such as edge states which are robust to disorder~\cite{wen,wen2}. Recently, engineered photonic systems have offered a new platform for studying topological features and topological band structure and their corresponding edge states have been observed in a number of experimental realizations ~\cite{wang,tiexp,rechtsman,topphoton}. More interestingly, a number of proposals have suggested that engineered photonic systems with strong two-body photon interactions and a $U(1)$ gauge field could realize Laughlin-type fractional quantum Hall states~\cite{cho,fraccav,nonequilibrium}. Parent Hamiltonians with three-body and higher order interactions can give rise to more exotic states like the Moore-Read state which is known to support non-Abelian excitations~\cite{threebody,kbody,stabilization,mooreread}. These systems potentially represent a solution to the quantum memory problem: non-Abelian quantum Hall states can store information that is robust to a wide variety of decoherence mechanisms~\cite{nayak}. There exists a rich phenomenology associated with these quantum liquids including excitations with fractional charge and exotic exchange statistics. The realization of parent Hamiltonians in photonic systems is thus of particular interest, especially since these systems can be realized in table-top devices and can be probed in novel ways.

Unlike the integer quantum Hall effect, the fractional quantum Hall effect requires repulsive interactions between particles. In the photonic setting, effective interactions among the photons would arise from engineered non-linearities in the system and are thus not universal~\cite{anderson}. These interactions will vary from site to site and for sufficiently strong disorder could give rise to attractive interactions. Sufficiently strong disorder in the interaction strength would destroy the quantum Hall ground state. Disordered repulsive interactions have been studied. It is particularly important to address the stability of a fractional quantum Hall liquid in the presence of sites in which the interactions among photons is attractive. Rather than considering an implementation-specific model, in this work we consider the effect of some density of these sites on the stability of the $\nu = 1/2$ Laughlin states through numerical simulations. We find that the quantum Hall liquid is robust to such defects, and our work gives tolerances for the strength and density of such defects associated with the relative strength of the disorder to the interaction strength.

We find three broad regimes which characterize the response of the liquid to these `interaction' defects. In the weak defect regime (for which first order perturbation theory holds), the wave function is largely unaffected by the defect. In the intermediate regime, the Laughlin state remains the ground state of the system, but the excited states are characterized by an increase in photon density around the defects. These regions have a characteristic length scale set by the magnetic length $\ell$. Finally, for sufficiently strong defects, pairs of photons co-localize around the defects and destroy the topological order. We find that particularly strong attractive interactions are required for this to occur.

The outline of this paper is as follow. In Section II, we discuss the bosonic Laughlin state and the characteristic energy scales in the liquid. Section III discusses the effects of an interaction impurity, detailing the physics of the three regimes discussed above. Finally, section IV presents our conclusions.

\section{Bosonic Laughlin States}

The FQHE is described by the filling factor $\nu$ which is defined as $\nu = N/N_\phi$ where $N$ is the number of particles and $N_\phi$ is the number of magnetic flux quanta (of strength $\Phi_0 = h/e$) piercing the system. For neutral particles, artificial magnetic fields can be synthesized (see~\cite{jaksch} for the case of atoms and~\cite{hafezi3} for photons). In an engineered lattice of cavity resonators, the dynamics of the photons is well-described by the Bose-Hubbard model~\cite{topphoton}. The hopping is given by
\begin{equation}
H_0 = - J \sum_{x,y} \hat{a}^\dagger_{x+1,y} \hat{a}^{\phantom\dagger}_{x,y} e^{-i \pi \alpha y} + \hat{a}^\dagger_{x,y+1} \hat{a}^{\phantom\dagger}_{x,y} e^{i \pi \alpha x} + \mbox{H.c.},
\label{eq:H0}
\end{equation}
where $\hat{a}_{x,y}$ annihilates a boson at site $(x,y)$. The parameter $\alpha$ is a $U(1)$ gauge term which mimics the effect of a magnetic field. For $\alpha = 0$, the hopping part of $H$ (which we will denote by $H_0$) [Eq.~(\ref{eq:H0})] gives rise to a band with a reduced mass $m^\ast = \hbar^2/2 J a^2$. For $\alpha \neq 0$, the spectrum of $H_0$ is a set of topological flat bands or Landau levels. The relation between $\alpha$ and $\ell$, the `magnetic' length which appears in (\ref{eq:laughlin}), is given by $\ell = a/\sqrt{2 \pi \alpha}$ where $a$ is the lattice spacing.

The continuum limit corresponds to $\alpha \ll 1$. In this limit, the cyclotron frequency is related to $\ell$ via
\begin{equation}
\omega_c = \frac{\hbar}{m^\ast \ell^2},
\label{eq:cyclotron}
\end{equation}
where $m^\ast$ is the reduced mass. The Landau level spacing is given by
\begin{equation}
\hbar \omega_c = 4 \pi \alpha J,
\label{eq:ell}
\end{equation}
valid for $\alpha \ll 1$.

On-site interactions are described by
\begin{equation}
H_{\rm int} = U_0 \sum_{x,y} \hat{n}_{x,y} \left( \hat{n}_{x,y} - 1 \right).
\end{equation}
Such interactions can be engineered in various systems\cite{cho,foss,sieberer,noh,deng,kasprazak,szymanska,byrnes,houck,nissen,raftery,barends,fitzpatrick,greentree,spillane,vetsch,tiecke,thompson,goban,hennessy,aoki,gorshkov,peyronel,carr,malossi}.
In the continuum limit $\alpha \ll 1$, the interacting system $H = H_0 + H_{\rm int}$ has a ground state which possesses a large overlap with the celebrated Laughlin wave function,
\begin{equation}
\Psi_m(z_1, z_2,...,z_N) = \prod_{j<k} \left(z_j - z_k \right)^m e^{-\sum_i|z_i|^2/4\ell^2},
\label{eq:laughlin}
\end{equation}
where $z_j = x_j + i y_j$ encodes the position of the $j^{\rm th}$ particle. For bosons, we have $\nu = 1/m$ with $m$ even.

In the absence of interactions (i.e. $H_{\rm int} = 0$), all the states associated with the lowest Landau level are degenerate. Interactions lift this degeneracy. In first order degenerate perturbation theory, valid for $U_0 \ll 4 \pi \alpha J$, states in the lowest Landau (the $i^{\rm th}$ state in the lowest Landau level is denoted $| \textrm{LLL}_i \rangle$) can then be diagonalize using $H_{\rm int}$ and have energies
\begin{equation}
E_i = \langle \textrm{LLL}_i | H_{\rm int} | \textrm{LLL}_i \rangle + \frac{1}{2} \hbar \omega_c,
\label{eq:deltan}
\end{equation}
where the index $i = 1,2,...$ enumerates the states in the lowest Landau level. For states on a finite torus, the first and second states are nearly degenerate ($E_1 \approx E_2$) and the many-body gap is
\begin{equation}
\Delta_0 = E_3 - E_1.
\end{equation}
From Eq.~(\ref{eq:deltan}), for $U_0 \ll J$, $\Delta_0$ is proportional to $U_0$. On the other hand, in the limit that $U_0 \gg J$, the hopping becomes the perturbation and thus $\Delta_0$ would be proportional to $J$. The low-lying spectrum for a simulation can be seen in Fig.~\ref{fig:redgapfit}. The low-lying many-body spectrum admits an interpretation as the creation of quasiparticles and quasiholes.

\section{Interaction Impurities}

\begin{figure}
\begin{center}
\includegraphics[bb=1 1 392 304, width = 9cm]{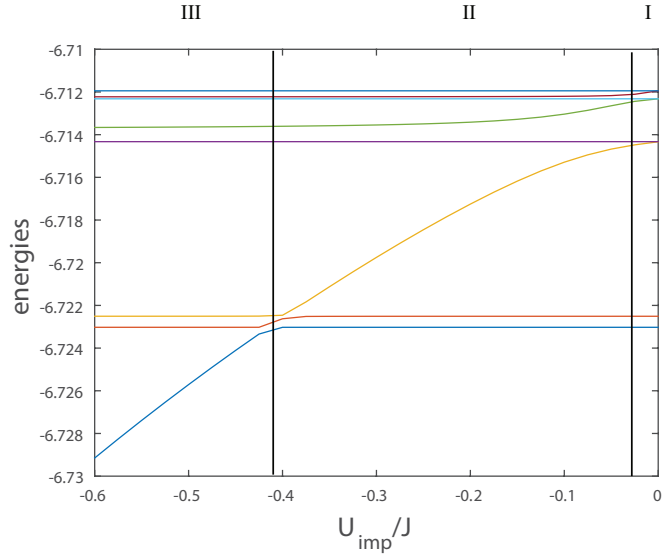}
\caption{Typical low energy spectrum of $H = H_0 + H_{\rm int}$ as a function of the interaction impurity $U_{\rm imp}$ in units of the hopping $J$. This simulation has $N = 2$ atoms with an $\alpha = 1/11$. The Landau level energy is given by $1.142J$, and $U_0 = 0.1J$. The system exhibits three regimes of interest: (I) the impurity interaction potential $H_{\rm imp}$ [Eq.~(\ref{eq:impurity})] represents a weak perturbation; (II) the photons tend to cluster around the impurity site(s); and (III) the localized state is the ground state of the system.}
\label{fig:redgapfit}
\end{center}
\end{figure}

We now turn to the main focus of the paper: describing the effects of a finite density of interacting defects. To simulate this, we numerically diagonalize $H = H_0 + H_{\rm int} + H_{\rm imp}$ on a variable size lattice (from $4 \times 4$ to $8 \times 8$) with periodic boundary conditions. The interaction impurities are described by
\begin{equation}
H_{\rm imp} = \sum_{(x,y) \in S} U_{\rm imp} \hat{n}_{x,y} \left( \hat{n}_{x,y} - 1 \right),
\label{eq:impurity}
\end{equation}
where $S$ contains either one or two sites, for instance $S = \{(1,1)\}$ or $\{(0,0),(4,4)\}$ so that the separation between two impurity sites is always much larger than $\ell$ and
\begin{equation}
n_{\rm imp} \ll \ell^{-2},
\end{equation}
where $n_{\rm imp}$ is the areal density of defects.

The full Hamiltonian is now given by $H = H_0 + H_{\rm int} + H_{\rm imp}$. We find that for $U_{\rm imp} > 0$, the spectrum remains relatively unchanged from the uniform case even when $U_{\rm imp}$ is comparable to $U_0$. For the remainder of the paper, we focus on $U_{\rm imp} < 0$. For $U_{\rm imp} = - U_0$, the impurity site(s) is rendered non-interacting. For $U_{\rm imp} < -U_0$, the on-site interactions are attractive, while for $U_{\rm imp} > - U_0$ they remain repulsive.

\begin{figure}
\begin{center}
\includegraphics[width=9cm]{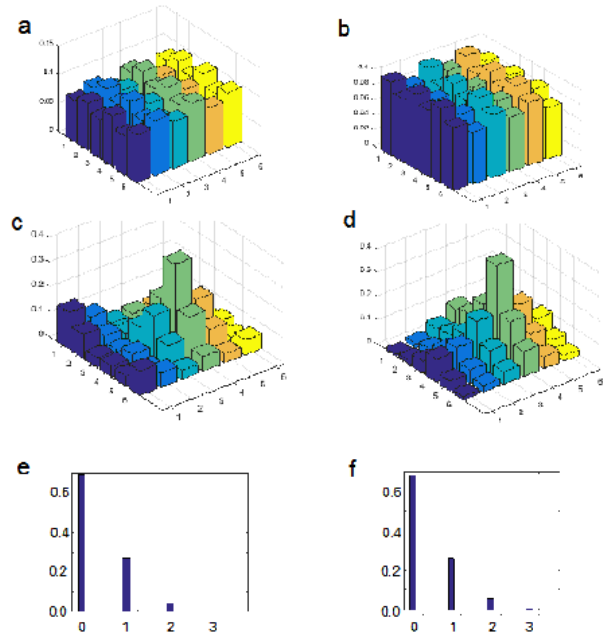}
\caption{(a)-(d) The average number of particles $\langle n_{(x,y)} \rangle$ on each site for the lowest four eigenvalues for an interaction impurity with $U/J = -1.6$ at site (4,4). (a)-(b) represent the ground state, while (c)-(d) show $\langle n_{(x,y)} \rangle$ for the first and second excited states. (e)-(f) show the probability of finding zero, one, two, or three particles at site $(4,4)$.}
\label{fig:number}
\end{center}
\end{figure}

The Laughlin wave functions remains an eigenstate of $H = H_0 + H_{\rm int} + H_{\rm imp}$ for all $U_{\rm imp}$ since $\langle \Psi_m | H_{\rm int} | \Psi_m \rangle = 0$. This follows from the fact that $\Psi_m$ [Eq.~(\ref{eq:laughlin})] vanishes whenever any two bosons are coincident ($z_j \rightarrow z_k$). In contrast, some excited states will be affected by the perturbation (see Fig.~\ref{fig:number}). One way to understand this is to interpret the many-body excited states in terms of quasiparticles and quasiholes: there is a finite amplitude for two bosons to be coincident in a region of length $\ell$ around a quasiparticle. Excited states with this property have energies that are functions of $U_{\rm imp}$.

In order to assess the stability of the ground state to these impurities, we will investigate the many-body gap $\Delta_0(U_{\rm imp})$ and its dependence on the density and strength of the impurity sites. The effect of the impurity site(s) is captured by the quantity
\begin{equation}
\Delta_i(U_{\rm imp}) = | \Delta_0(U_{\rm imp}) - \Delta_0(0) |.
\end{equation}
In order to address the physics of the interaction impurity, we consider three regimes which we denote the perturbative, localized, and strongly attractive regimes as described below. We will employ both the scaling behavior of $\Delta_i(U_{\rm imp})$ and the Chern number to characterize these regimes. For some critical $U_{\rm imp}$, we expect that $\Delta_{i}$ will vanish, signaling a phase transition in the system.

\begin{figure}
\begin{center}
\includegraphics[width=9cm]{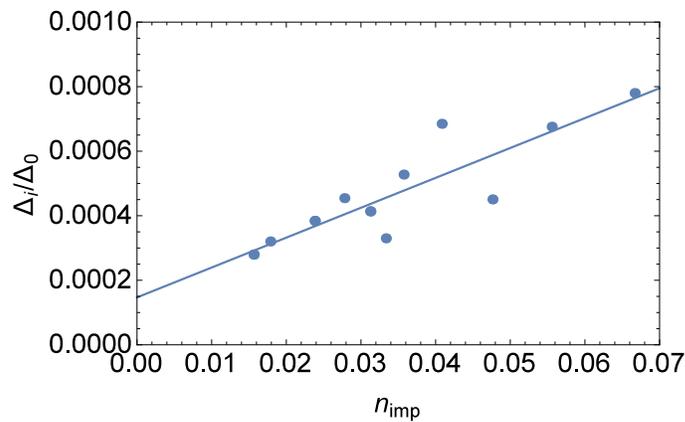}
\caption{The reduction ($\Delta_i$) of the many-body gap in the perturbative regime, with $U_{\rm imp}/U_0 = 0.01$. The data shows good agreement with Eq.~(\ref{eq:leastsquares}). The data shown includes simulations with $2$ to $5$ particles with lattices in size from $4 \times 4$ to $9 \times 9$. In all simulations, $\nu = \frac{1}{2}$.}
\label{fig:small}
\end{center}
\end{figure}

The perturbative regime is characterized by impurities with $|U_{\rm imp}| \ll U_0$; this is region $I$ as shown in Fig.~\ref{fig:redgapfit}. In this limit, (regardless of the the sign of $U_{\rm imp}$), first order perturbation theory applies and the first excited state is essentially unchanged from the clean case. In the limit in which $4\pi J \alpha \gg U_{0}$, each term in $H_{\rm int}$ can be treated using first order perturbation theory. Since the unperturbed state is uniform, each term contributes equally to the gap. This implies that $\Delta_i \propto a^2 n_{\rm imp} \left| U_{\rm imp }\right|$. Since $\Delta_0 \propto U_0$ ($\Delta_0$ is the gap in the absence of any impurities), we obtain
\begin{equation}
\frac{\Delta_i}{\Delta_0} \approx a^2 n_{\rm imp} \frac{\left|U_{\rm imp}\right|}{U_0}.
\label{eq:Delta1}
\end{equation}
The results of our simulations in the perturbative regime are shown in Figs.~\ref{fig:small}. The least squares fit shown in (a) is given by
\begin{equation}
\Delta_i/\Delta_0 \approx 1.5 \times 10^{-3} + 9.3 \times 10^{-3} n_{\rm imp}.
\label{eq:leastsquares}
\end{equation}
This is in good agreement with Eq.~(\ref{eq:Delta1}) with $U_{\rm imp}/U_0 = 0.01$ for the simulations shown.

\begin{figure}
\begin{center}
\includegraphics[width = 9cm]{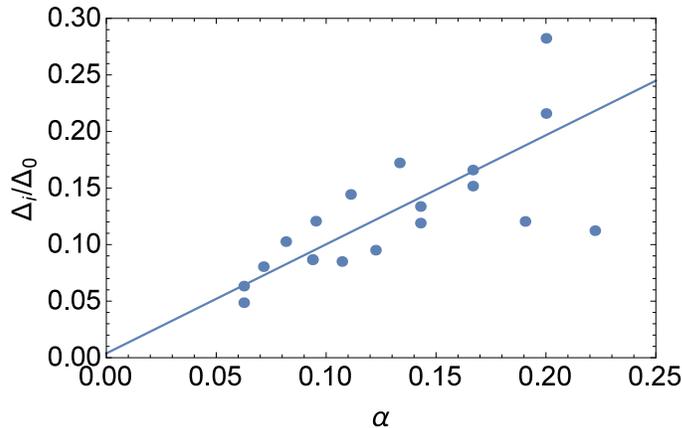}
\caption{Values of $\Delta_i / \Delta_0$ for $U_{\rm imp} = - U_0$ plotted as a function of $\alpha$. The approximate linearity of the data (plotted as a function of $\alpha$) is an indication of the localization of the excited state around the impurity/impurities.}
\label{fig:large}
\end{center}
\end{figure}

The non-perturbative regime is characterized by $U_{\rm imp} \sim - U_0$, shown as region $II$ in Fig.~\ref{fig:redgapfit}. In this regime, there is a tendency for photons to become localized around the impurity sites. However, due to hopping this localization is imperfect and instead the bosons are localized to a region of characteristic size $\ell$ around the impurity sites. This feature may also be understood in the context of the \emph{plasma analogy}. Laughlin observed that the wave function (\ref{eq:laughlin}) has a charge density which is related to a collection of interacting line charges in 2D in the presence of a uniform background charge~\cite{plasma}. In this picture, particles will cluster around an energetically favorable region or site with a healing length $\sim \ell$.

Thus, in this regime (and for low particle density) we expect that bosonic particle density is localized to an approximate area $\pi \ell^2$ around each defect. To the extent that this localization is perfect and the wave function is uniform in this region, the ratio $\Delta_i / \Delta_0$ would be given by the fraction of the wave function which covers an impurity site, namely $a^2 / \pi \ell^2 = 2 \pi \alpha$. For the regime in which $U_0 \ll 4 \pi \alpha J$ (and thus $\Delta_0 \propto U_0$), we have that $\Delta_i \propto (a^2/\ell^2) U_{\rm imp}$ and thus
\begin{equation}
\Delta_i \propto \alpha U_{\rm imp}.
\label{eq:Delta2}
\end{equation}
We have tested this relationship for a broad parameter regime and find that it holds outside of the perturbative regime. A test of this behavior, shown in Fig.~\ref{fig:large}, validates this localization picture. Moreover, this behavior is distinct from the behavior predicted by Eq.~(\ref{eq:Delta1}). The scaling exhibited in Fig.~\ref{fig:large} is only approximate, and deviations are expected. First, the scaling relation~(\ref{eq:Delta2}) assumes that the wave function is uniform in an area $\sim \ell^2$ centered on the impurity/impurities. This is an approximation, and the correlations in $\langle n_{(x,y)} \rangle$ and $\langle n_{(x,y)}^2 \rangle$ will vary in this region [see Fig.~\ref{fig:number} (c-f)]. For higher densities of particles, screening of the impurity site may occur. We note that higher particle density corresponds to large $\alpha$ and this is the region in which deviations from Eq.~(\ref{eq:Delta2}) are largest. For $\alpha \approx 1/4$, the continuum approximation breaks down suggesting that lattice effects may also play a role.

\subsection{Strongly Attractive Regime}

\begin{figure}
\begin{center}
\includegraphics[width=9cm]{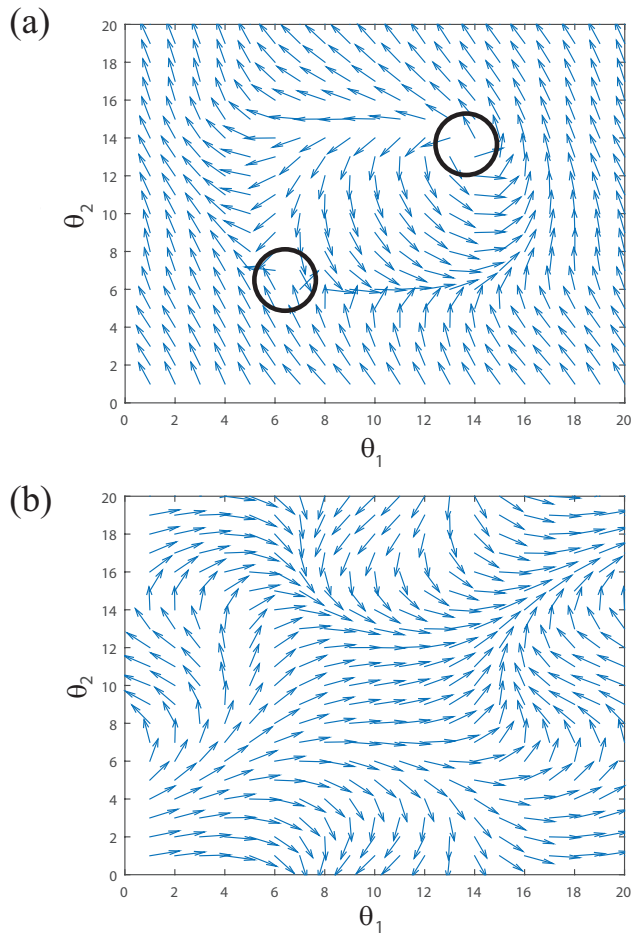}
\caption{Plot of $\Omega(\theta_1,\theta_2)$ for (a) the Laughlin state for $U_{\rm imp} \approx 0$ and (b) the non-Laughlin ground state for $U_{\rm imp} \approx - 3 U_0$ (see Ref.~\cite{fqhebh} for a definition). Vortices in $\Omega(\theta_1,\theta_2)$ indicate a nonzero Chern number.}
\label{fig:omega}
\end{center}
\end{figure}

For sufficiently strong impurities, a level crossing occurs; this regime is indicated by $III$ in Fig.~\ref{fig:redgapfit}. The nature of the wave functions of the old ground states do not change: the overlap with the Laughlin state remains very close to 1 (within 1\%). It is well known that Laughlin state for $\nu = 1/2$ is characterized by a Chern number of 1~\cite{fqhebh}. We now address whether the transition seen in Fig.~\ref{fig:redgapfit} represents a topological phase transition. It is quite likely that the new ground state is topologically trivial given that it is non-degenerate. To confirm this, we employ the method developed by Hatsugai~\cite{hatsugai} and Kohmoto~\cite{kohmoto} to calculate the Chern number of the new ground state. The Chern number can be related to the vorticity of a quantity $\Omega(\theta_1, \theta_2)$, where $\theta_{1,2}$ are twist angles associated with the toroidal boundary conditions~\cite{fqhebh,hatsugai,kohmoto}. The method is involved and requires fixing the gauge of the wave function. In Fig.~\ref{fig:omega}, we have plotted $\Omega(\theta_1,\theta_2)$ for the Laughlin ground state and the non-degenerate ground state. As shown in Fig.~\ref{fig:omega}(b), the latter possesses no vorticity. This indicates that the Chern number is zero and thus the new ground state is topologically trivial.

In a best case scenario for the realization of a Laughlin liquid in an actual experiment, it's clear that the localized regime is to be avoided if possible. However, even in this case, it may still be possible to access the Laughlin liquid. If an injected photon has minimal overlap with a localized ground state, there may be a large amplitude for the photon to be in the Laughlin state and thus realize fractional quantum Hall physics. Based on our understanding of the localized regime, this may be accomplished if the photon is injected into the system far $(\gg \ell)$ from the defective sites.

\section{Conclusions and Outlook}

We have studied the role that interacting impurities play in lattice realizations of the fractional quantum Hall effect. We have outlined three different regimes which characterize the response of the topological liquid. Our findings point to the robustness of the Laughlin liquid to impurities of these type. Only for impurity sites which host very strong attractive interactions does the system undergo a topological phase transition to a trivial phase. These findings are an important feasibility consideration for the realization of photonic quantum Hall liquids and bode well for their creation.

\section{Acknowledgements}

We thank Alexey Gorshkov and Ignacio Cirac for conversations about this work. We are also grateful to Tobias Grass and Guanyu Zhu who offered numerous suggestions for improving the manuscript. We also would like to thank Brandon M. Anderson for useful discussions. This research was supported under National Science Foundation PFC at the Joint Quantum Institute, and ARO-MURI, AFOSR- MURI FA95501610323, Sloan Fellowship, YIP-ONR.

\end{document}